 \documentclass[reprint,aps,superscriptaddress,twocolumn]{revtex4-1}

\usepackage[utf8]{inputenc}  
\usepackage[T1]{fontenc}     
\usepackage[british]{babel}  
\usepackage[sc,osf]{mathpazo}
\usepackage[scaled=0.86]{berasans}  
\usepackage{graphicx} 
\usepackage[babel]{microtype}  
\usepackage{amsmath,amssymb,amsthm,bm,amsfonts,mathrsfs,bbm} 

\usepackage{xspace}  
\usepackage{pgf,tikz}
\usepackage{xcolor}
\usepackage{multirow}
\usepackage{array}
\usepackage{bigstrut}
\usepackage{braket}
\usepackage{color}
\usepackage{natbib}
\usepackage{multirow}
\usepackage{mathtools}
\usepackage{float}
\usepackage{xcolor,colortbl}
\usepackage{color}
\usepackage[justification=justified, format=plain]{subcaption}
\usepackage[justification=raggedright]{caption}
\usepackage[normalem]{ulem}
\usepackage{etoolbox}
\usepackage{xparse}
\usepackage{cleveref}
\usepackage{crossreftools}
\usepackage{graphicx}
\usepackage{comment}
\usepackage{bm}
\usepackage{mathtools}

\newcommand\numberthis{\addtocounter{equation}{1}\tag{\theequation}}

\newcommand{\cI}{\mathcal{I}}

\newcommand{\cC}{\mathcal{C}}
\newcommand{\cD}{\mathcal{D}}

\newcommand{\cH}{\mathcal{H}}
\newcommand{\cN}{\mathcal{N}}
\newcommand{\cP}{\mathcal{P}}
\newcommand{\cQ}{\mathcal{Q}}
\newcommand{\cR}{\mathcal{R}}

\newcommand{\cT}{\mathcal{T}}
\newcommand{\cX}{\mathcal{X}}
\newcommand{\cY}{\mathcal{Y}}
\newcommand{\cZ}{\mathcal{Z}}

\newcommand{\indep}{\perp \!\!\! \perp}

\theoremstyle{plain}
\newtheorem{theorem}{Theorem}

\newtheorem{lemma}[theorem]{Lemma}

\newtheorem{proposition}[theorem]{Proposition}
\newtheorem{definition}[theorem]{Definition}

\newtheorem{fact}[theorem]{Fact}
\theoremstyle{definition}
\newtheorem{remark}[theorem]{Remark}

\newtheorem{example}[theorem]{Example}

\renewcommand{\epsilon}{\varepsilon}

\DeclareMathOperator{\Tr}{Tr}

\newcommand{\ketbra}[1]{\ket{#1}\!\!\bra{#1}}

\newcommand{\one}{\leavevmode\hbox{\small1\kern-3.8pt\normalsize1}}

\begin{document}
\title{Noise is resource-contextual in quantum communication}

\author{Aditya Nema}
\affiliation{Graduate School of Informatics, Nagoya University, Japan}

\author{Ananda G. Maity}
\affiliation{Networked Quantum Devices Unit, Okinawa Institute of Science and Technology Graduate University, Onna-son, Okinawa 904-0495, Japan}

\author{Sergii Strelchuk}
\affiliation{Department of Applied Mathematics and Theoretical Physics, University of Cambridge, Cambridge CB30WA, UK}

\author{David Elkouss}
\affiliation{Networked Quantum Devices Unit, Okinawa Institute of Science and Technology Graduate University, Onna-son, Okinawa 904-0495, Japan}
\maketitle
{\bf Estimating the information transmission capability of a quantum channel remains one of the fundamental problems in quantum information processing. In contrast to classical channels, the information-carrying capability of quantum channels is contextual. One of the most significant manifestations of this is the superadditivity of the channel capacity: the capacity of two quantum channels used together can be larger than the sum of the individual capacities. Superadditive behaviour helps to overcome noise and dramatically improve communication rates. We uncover a stark counterintuitive behaviour of noise: we introduce a one-parameter family of channels for which as the parameter increases its (less resourceful) one-way quantum and private capacities increase while its (more resourceful) two-way capacities decrease. Our constructions demonstrate that noise is context dependent in quantum communication. We also exhibit a one-parameter family of states with analogous behavior with respect to the one- and two-way distillable entanglement and secret key highlighting new properties of quantum resources.   
 }

\section*{Introduction}
Determining the capability of a quantum channel for sending quantum information is a notoriously difficult problem. Classical communication channels can be effectively characterized by a single number -- capacity -- which completely describes its ability to convey information. The analogous expression for quantum channels is easy to write down, but exceptionally difficult to compute in general. It involves regularization over an unbounded number of uses of the channel, making the task of characterizing the potential for sending quantum information computationally intractable. 

Di Vincenzo {\it et al.}~\cite{divincenzo1998quantum} were the first to observe that regularization is necessary by showing the underlying entropic quantity, the coherent information to be superadditive. Since then, there has been a major effort to understand superadditivity and its relation to the computation of capacity in the context of sending quantum information but also for other communication tasks where capacity is given by a regularized formula~\cite{wilde2013quantum,zhu2018superadditivity,zhu2017superadditivity, Elkouss2015private,smith2009extensive,brandao2012does,shirokov2015superactivation,leditzky2022generic,Manik_superadditivity,leung2014maximal,li2009private,singh2023simultaneous}. 
Notably, since generally regularized formulas are the only proxy for capacity, even elementary questions such as whether or not a channel has positive capacity have no known algorithmic answer~\cite{cubitt2015unbounded} (see \cite{siddhu2021entropic,singh2022detecting} for recent progress). 
On the other hand, there exist a few families of channels, such as degradable channels \cite{Devetak_dephasing}, for which both the capacity for sending quantum information and private classical information are given by the coherent information. 

The capacity of quantum channels can itself be superadditive. This property implies that the utility of a channel may depend on the other accompanying channels. Winter {\it et al.}~\cite{winter2016potential} defined the potential capacity of a channel as its maximum information-carrying capability when used in combination with an auxiliary contextual channel. Quantum capacity is highly contextual: it is known that there exist channels with zero quantum capacity but positive potential capacity~\cite{yard2011quantum}. 

Moreover, in sharp contrast to the classical case, there are many new communication protocols enhanced by the presence of different resources. For each of them, we wish to determine the optimal communication rate, giving rise to a different channel capacity. 

Here, we consider two tasks: the transmission of quantum and private classical information. We consider the resources of free classical communication either from sender to receiver or bi-directional. Depending on the type of information transmitted -- quantum information or private classical information --  we will refer to the corresponding capacities as one- and two-way quantum and private capacities.

While these two capacities are quantitatively different, intuitively one would expect them to have the same qualitative behavior in the presence of noise. The erasure channel, for which we have full characterization, supports this intuition. The erasure channel transmits the input with probability $1-\lambda$ and outputs an erasure flag with probability $\lambda$. It has one-way capacity proportional to $1-2\lambda$ \cite{bennett1997capacities} and two-way capacity proportional to $1-\lambda$ \cite{bennett1997capacities} (see also \cite{entanglemententropy_capacity}). In other words, the two-way capacity is quantitatively larger but has the same monotonic behavior with respect to $\lambda$.

\begin{figure}[h]
    \centering
    \includegraphics[width=8.3cm]{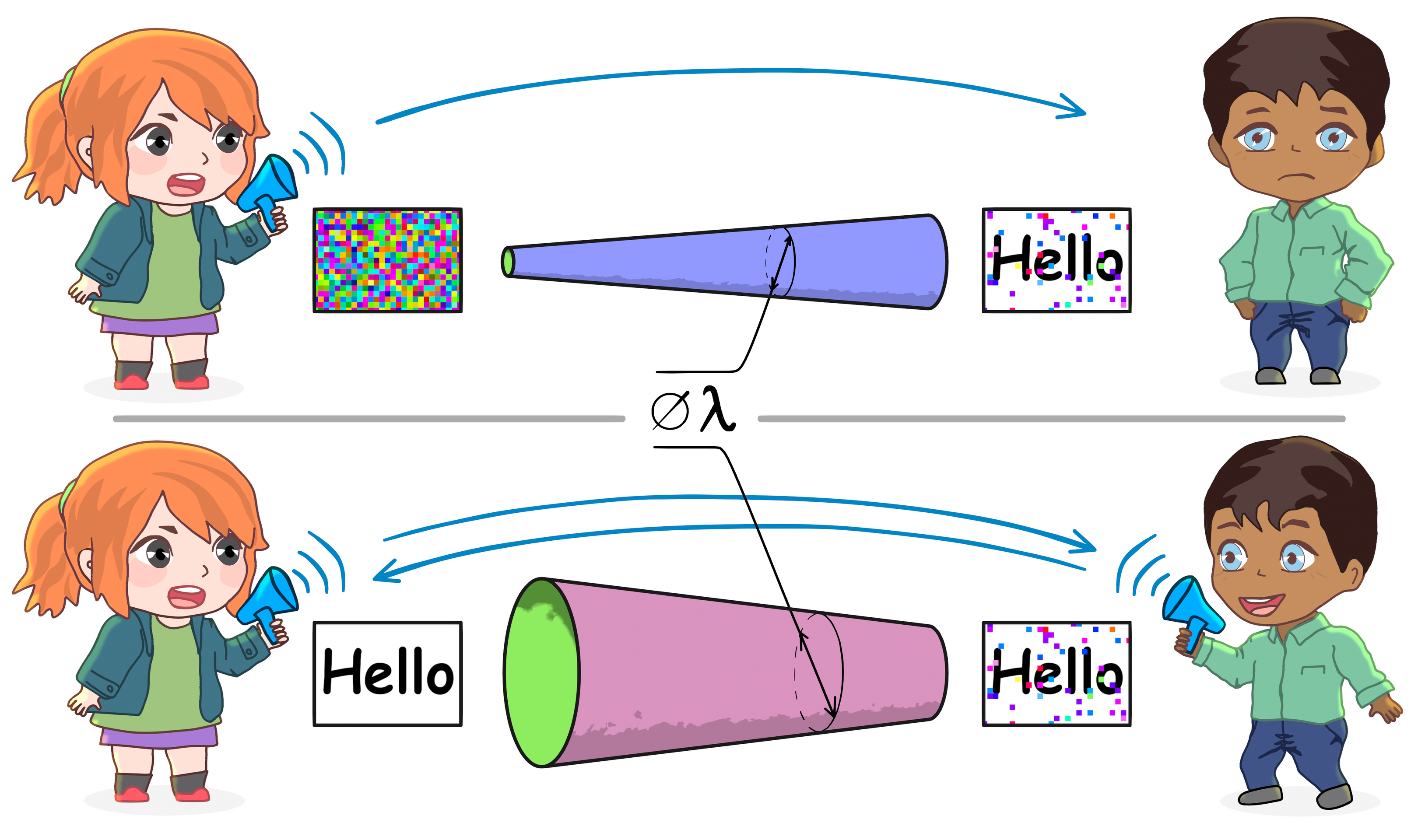}
        \caption{Noise resource-contextuality: capacity of a family of channels for transmitting quantum information as a function of the noise parameter $\lambda$. The top scenario corresponds to the capacity in the absence of feedback (one-way capacity) and the bottom scenario corresponds to the capacity with feedback (two-way capacity). Intuitively, one would expect both capacities to be quantitatively different but have similar qualitative behavior; that is, either both increase or both decrease as a function of the noise parameter. Here we show that the very meaning of noise can depend on the resources. We exhibit families of channels for which as the two-way capacity decreases, the one-way capacity increases. Increasing $\lambda$ represents noise for two-way communications while decreasing $\lambda$ represents noise for one-way communications. }
    \label{fig:pop_summary}
\end{figure}
In this work, we show that contrary to the intuition, noise is resource-contextual. We present a one-parameter family of degradable channels, i.e. for which the one-way quantum and private capacities are fully characterized by the coherent information. An increase of the parameter leads to the decrease in the two-way quantum capacity, and thus can be regarded as `noise'. Surprisingly, it has the opposite effect on the one-way communication rates: the one-way capacity increases. Beyond degradable channels, we prove a weaker version of this phenomenon for a family of channels with superadditive coherent information. Achieving this effect is not possible for classical communication over classical noisy channels. However, we identify an analogous behaviour in the scenario of private communication over classical wiretap channels.\\
\section*{Results}

We will present our results in finite dimensional Hilbert spaces denoted by $\cH$ with $\text{dim}(\cH)=|\cH|$, $\varphi$ and $\pi^A$ denote a pure state of the form $\ketbra{\varphi}$ and completely mixed state $\frac{I^A}{|A|}$ respectively. $H(\rho)$ denotes the von Neumann entropy of the state $\rho$ and is defined as $H(\rho):=-\Tr(\rho \log \rho)$. We also use $H$ for the Shannon entropy when it is clear from the context. Given $\{p_i\}_{i=1}^n,~\sum_{i=1}^np_{i}=1,~p_i\geq0$, $H(\{p_1,\ldots,p_n\}):=-\sum_{i=1}^np_i\log p_i$ and if $n=2$, $H(p):=H(\{p,1-p\})$. The short-hand notation, $\mathrm{diag}[a_1,\;a_2, \ldots]$ denotes a diagonal matrix with the diagonal elements $a_1,\;a_2, \ldots$ whereas random variables are denoted by plain capital letters $X,Y,Z$. The operation `$\bigoplus$' between two channels represents the direct sum of the range space of the underlying channels. 
The one-way quantum capacity $\cQ(\cN)$ of the channel $\cN$, defined as the maximum number of qubits that can be reliably transmitted per channel use given free access to forward classical communication. The `coherent information' $I_c(\cN)$ of $\cN$ and is defined as 
$I_c(\cN) := \max_{\ket{\phi}^{AA'}}I_c(\cN, \rho),$
where $I_c(\cN, \rho):= H(B)_\omega - H(AB)_\omega$, 
$\omega^{AB}:= (\cI^A \otimes \cN^{A' \to B}) \ketbra{\phi^{AA'}}$ 
and $\ket{\phi^{AA'}}$ is a state such that $\rho^A=\Tr_{A'}(\ketbra{\phi^{AA'}})$.
$\cQ(\cN)$ is lower bounded by the coherent information $I_c(\cN)$~\cite{lloyd1997capacity,devetak2005private}. While in general it is a strict lower bound, for a degradable channel, it fully characterises its capacity i.e,
$\cQ(\cN)= I_c(\cN)$~\cite{Cubitt_Degradability}.
Similarly, $\cQ_\leftrightarrow(\cN)$ represents the capacity with free classical communication allowed at the transmitter (classical feedforward) as well as the receiver end (classical feedback), henceforth referred as the two-way quantum capacity. 
The private classical capacity $\cP(\cN)$ of $\cN$ quantifies the maximum rate at which the users of the channel can communicate classical information privately, and analogously $\cP_\leftrightarrow(\cN)$ denotes the private capacity with two-way (public) classical communication assistance. We refer to \cite{wilde2013quantum} for a rigorous treatment of these notions.

{\bf Channels with increasing one-way capacity and decreasing two-way capacity.--} The main building blocks of our construction are given by a family of channels denoted by $\cN_{\lambda,\; p}$ that are parameterized by the parameters $\lambda, p$. Later we will set $p$ as a function of $\lambda$ or vice-versa singling out the family of channels $\cN_{\lambda, \; p}$ depending on a single parameter. 

\begin{figure}[h]
    \centering
    \includegraphics[width=8.3cm]{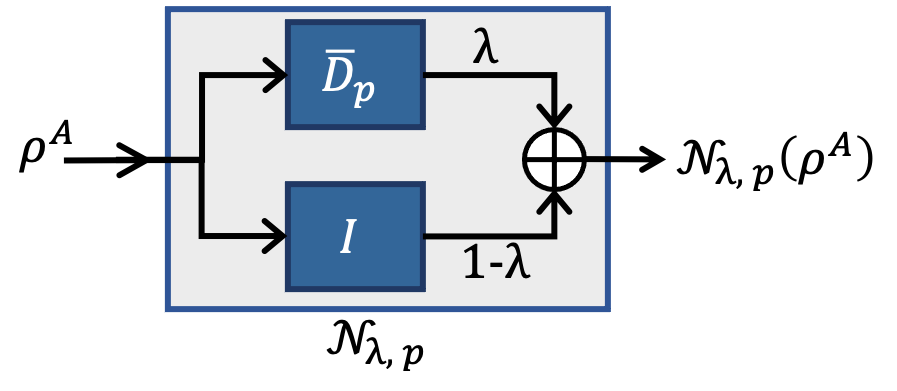}
        \caption{Weighted direct sum construction. $\overline{\cD_p}$ is the complement of a qubit dephasing channel with parameter $p$, $\cI$ is the noiseless channel, and $\lambda,\;p \in [0,1]$. This way of combining quantum channels is also referred to as `gluing'~\cite{Vikesh_superadditivity}.}
    \label{fig:qubit_channel}
\end{figure}

The channel construction is depicted in Fig.~\ref{fig:qubit_channel}. It maps states from the input space $A$ to space $B$ with $|\cH_A|=2$, $|\cH_B|=4$ and $\cH_C$ represents the environment.
Let $\ket{\varphi_0}:= \sqrt{1-p} \ket{0} + \sqrt{p} \ket{1}$, and $\ket{\varphi_1}:= \sqrt{1-p} \ket{0} - \sqrt{p} \ket{1}$.
The action of the channel and its complement on an arbitrary input state can be described as:
\begin{flalign} \label{eq:N(rho)}
\cN_{\lambda, \; p}(\rho):= (1-\lambda) \rho \oplus \lambda \overline{\cD_p}(\rho),
\end{flalign}
and
\begin{equation} \label{eq:complement_N}
    \overline{\cN_{\lambda, \; p}}(\rho):= (1-\lambda)  \ketbra{f} \oplus \lambda \cD_p(\rho)
\end{equation}
where $\ket{f}$ is a fixed pure state and $\cD_p,\overline{\cD_p}$ denote respectively the dephasing channel and its complementary. These channels are defined as: $\cD_p(\rho):=(1-p)\rho +p Z \rho Z$ and $\overline{\cD_p}(\rho):=\bra{0}\rho\ket{0} \ketbra{\varphi_0} + \bra{1} \rho \ket{1} \ketbra{\varphi_1}$. 

The channel in Equation~\ref{eq:complement_N} is the so-called `dephrasure' channel \cite{leditzky_dephrasure}. The dephrasure channel and its complementary channel (given in Equation~\ref{eq:N(rho)}) have proved to be a fertile ground for the study of coherent information superadditivity \cite{leditzky_dephrasure, Experimental_superadditivity20, siddhu2021entropic, Vikesh_superadditivity, bausch2020quantum}.

One can easily check that the output entropy of $\cN_{\lambda, \; p}$ and $ \overline{\cN_{\lambda, \; p}}$ are respectively given by:
\begin{align*} \label{eq:entropy_N}
    H(\cN_{\lambda, \; p}(\rho))=H(\lambda) + \lambda H(\overline{\cD_p}(\rho)) + (1-\lambda) H(\rho), \numberthis
\end{align*}
and 
\begin{align*} \label{eq:entropy_complement_N}
    H(\overline{\cN_{\lambda, \; p}}(\rho))&=H(\lambda) + \lambda H(\cD_p(\rho)) + (1-\lambda) H(\ketbra{f})\\
    &= H(\lambda) + \lambda H(\cD_p(\rho)). \numberthis 
\end{align*}
The channel $\cN_{\lambda, \; p}$ from Equation~\eqref{eq:N(rho)} has several important properties. First, its coherent information is invariant with respect to conjugation of the input state by a qubit Pauli $X$ or $Z$. Second, for $\lambda \in [0,1/2]$,  $\cN_{\lambda, \; p}$ is degradable. Proofs of these facts are found in SI~\ref{prop:Icoh_invariance} and \ref{prop:degraded_channel}. We refer to \cite{leditzky_dephrasure} for the original proofs and additional analysis of the channel.

The simple structure of $\cN_{\lambda, \; p}$ allows us to characterize its capacity. In particular, for $\lambda \in [0,1/2]$:
$\cP(\cN_{\lambda, \; p})=\cQ(\cN_{\lambda, \; p})=1-\lambda(2-H(p))$, and for $\lambda \in [0,1]$: ${\cP}_\leftrightarrow(\cN_{\lambda, \; p})={\cQ}_\leftrightarrow(\cN_{\lambda, \; p})=1-\lambda$.

Combining the above and choosing an appropriate relation between $p$ and $\lambda$, we construct a one-parameter family of channels $\{\cN_x\}_{x\in[a,b]}$ for which ${\cQ}_\leftrightarrow(\cN_x)>{\cQ}_\leftrightarrow(\cN_y)$ and ${\cQ}(\cN_x)<{\cQ}(\cN_y)$ for $x<y$ and $x,y\in[a,b]$ (the same statement holds for the private capacity). We illustrate this effect with two examples. First, pick $p(\lambda)=4\lambda-1$ which leads to the desired behavior in the range $\lambda\in[0.25,0.3125]$ (see  Figure~\ref{fig:quantum_capacity}). Second, select $\lambda(p)=p/\log p$ to observe the stated behavior in the range $p\in[0.35,0.5]$ (see  Figure~\ref{fig:quantum_capacity_p}). Additionally with this second parametrization the one-way capacity reaches its maximum for $p=1/2$, matching the two-way assisted capacity (see SI~\ref{Appendix:comp_deg} for details). 

\section*{Discussion}
Capacities of a noisy quantum channel exhibit puzzling behavior like superadditivity~\cite{divincenzo1998quantum,li2009private}, super-activation~\cite{yard2011quantum} and non-convexity~\cite{yard2011quantum,elkouss2016nonconvexity}. It can be linked to the contextuality of channel capacity \cite{winter2016potential}. 
Here, we find that the notion of noise must also be considered relative to the communication context. If we associate the noise to a continuous reduction of transmission capabilities for a fixed task and resource set (for instance quantum communication assisted by two-way classical communication) then we can find that for the same task and different auxiliary resources, the behavior is reversed. 

The proofs of superactivation, superaddivity and non-convexity of quantum and private capacity rely on the superadditivity of the coherent or private information. Remarkably, one of the two families of channels that we present here has additive coherent information, both the channels themselves as well as their complementary channels, discarding the possibility that the noise resource-contextuality is a byproduct of coherent information superadditivity.

Noise resource-contextuality is not limited to the complement of antidegradable dephrasure channels; Proposition~\ref{prop:discretefamily} gives a sufficient condition to find an infinite discrete family of channels for which the lower and upper bounds on the one-way assisted quantum capacity increases and the two-way assisted capacity is continuously decreasing. 

For conciseness, we give an example of Proposition~\ref{prop:discretefamily} with a family of non-degradable complementary of dephrasure channels. However, it is simple to construct other examples. Our characterization of two-way capacity holds for all direct sum constructions of identity and entanglement breaking channel. These constructions, corresponding to the complementary of generalized erasure channels \cite{Vikesh_superadditivity}, can be leveraged to construct new explicit examples. The requirement is to find a parametrization such that the one-way capacity starts at zero and a lower bound on the one-way capacity that is increasing in the parameter of the antidegradable channel family. This suffices, since one can then construct an upper bound from continuity arguments that is also increasing.  

Our result strengthens the difference between classical information theory, where channel capacity represents complete information about its utility and quantum information theory, where the utility of a channel strongly depends on the context. Indeed, the behavior we describe in our findings cannot exist classically.  However, going beyond unicast communication scenarios, we identify the curious case of a similar behaviour for classical wiretap channels. 

Lastly, our construction is simple enough to be experimentally feasible with present-day technology. Yu et al.  \cite{Experimental_superadditivity20}, while investigating the coherent information for a `dephrasure channel' in an optical set-up, consider a dephasing channel, an erasure channel as well as their complementary channels. In order to realize our channel, one only needs access to the complementary of the dephrasure channel and the capability of fine-tuning the dephasing parameter $p$ and the erasure probability $\lambda$.

\section*{Methods}
{\bf Proof sketch of the capacity characterization.} 
For characterizing the one-way capacities $\cP$, it is helpful to note that $\cN_{\lambda, \; p}$ is degradable (see SI~\ref{prop:degraded_channel} for the proof) and hence its quantum and private capacities coincide \cite{Degradable_capacity}. Moreover, the quantum (and private) capacity is then given by the coherent information (by \cite[Theorem~2]{Degradable_capacity}, also \cite{Devetak_dephasing}). For a detailed argument see SI \ref{proof_theorem1}.

For characterizing the two-way capacities ${\cP}_\leftrightarrow$, one may note that $\cN_{\lambda, \; p}$ is teleportation stretchable \cite{entanglemententropy_capacity}, (also from \cite[Theorem~12]{Er_converse_Wilde}). Thus the two-way classically assisted private capacity and hence the quantum capacity is upper bounded by the relative entropy of entanglement of $\cN_{\lambda, \; p}$ (formally defined in Section~\ref{sec:entropyupperbound} of the SI) which evaluates to $1-\lambda$. The upper bound is achievable and can be observed by inspection; the encoder can send half of a maximally entangled state. This procedure prepares a joint maximally entangled state when the channel acts as the identity which occurs with probability $1-\lambda$. Exploiting the direct-sum structure of the channel, the communicating parties can distinguish between the action of $\overline{\cD_p}$ and $I$ and consume the maximally entangled states to communicate noiselessly at a rate of $1-\lambda$ (see SI \ref{proof_theorem1}). 

\begin{figure}[h]
    \centering
    \includegraphics[width=8.3cm]{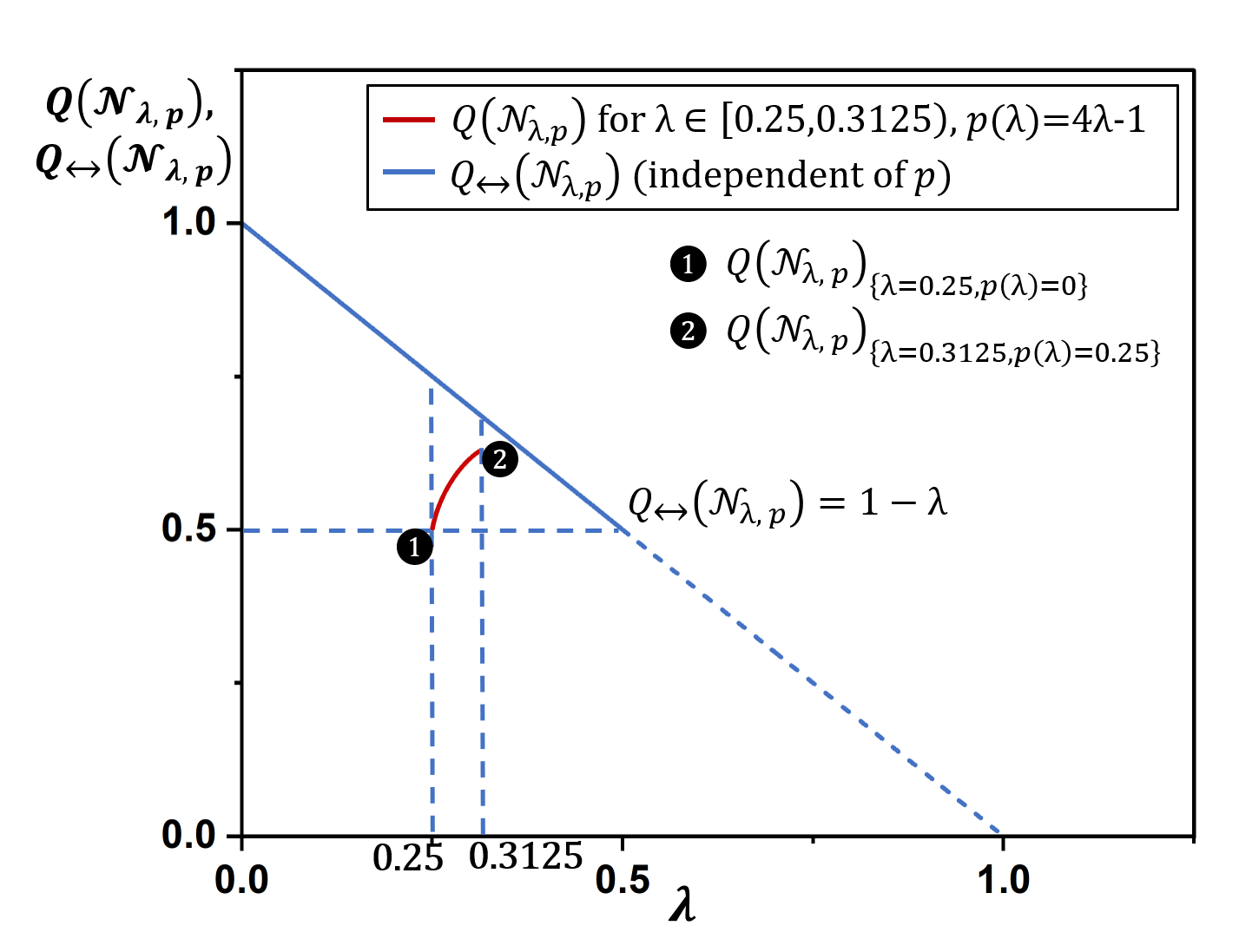}
        \caption{One- vs two-way capacity of $\cN_{\lambda, \; p}$ as a function of $\lambda$ when $p(\lambda)=4\lambda-1$. In the range  $\lambda\in[0.25,0.3125]$ the one-way quantum (and private) capacity monotonically increases while the two-way quantum (and private) capacity decreases. 
        }
    \label{fig:quantum_capacity}
\end{figure}

\begin{figure}[h]
    \centering
    \includegraphics[width=8.3cm]{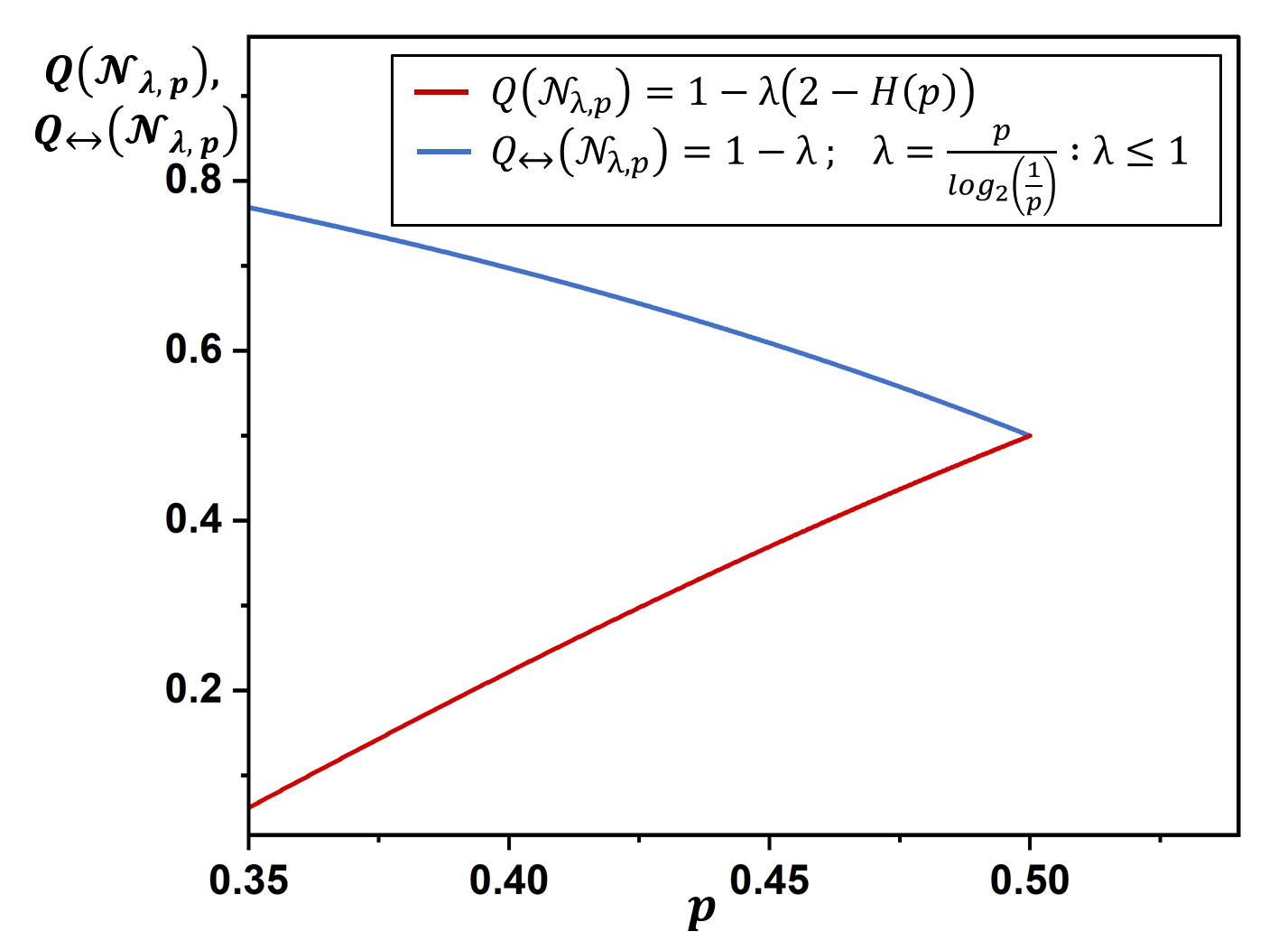}
        \caption{ 
        One- vs two-way quantum (and private) capacity of $\cN_{\lambda, \; p}$ as a function of $p$ when $\lambda(p)=p/\log p$. In the range  $p\in[0.35,0.5]$ the one-way quantum (and private) capacity monotonically increases while the two-way quantum (and private) capacity decreases.}
    \label{fig:quantum_capacity_p}
\end{figure}

{\bf  One-way and two-way distillable entanglement and distillable secret key.} The previous discussion can directly be extended to the distillable entanglement and the distillable secret key \cite{Devetak_Winter}. We consider the family of bipartite states given by the Choi states of the above channels and calculate its two-way capacities by evaluating the two-way distillable entanglement of the Choi state (see SI \ref{proof_theorem1}). In particular, distillable entanglement 
and secret key are upper bounded by the relative entropy of entanglement \cite{Horodecki_key}.
The achievability of this bound is shown in SI \ref{proof_theorem1}.

The one-way capacity coincides with the coherent information of the Choi state (see SI \ref{proof_theorem1}). In principle, the coherent information is only a lower bound on the one-way distillable entanglement and secret key of a state. However, for degradable states the coherent information coincides with the distillable entanglement \cite{leditzky2017useful} and also the distillable key. This can be straightforwardly deduced from \cite[Equation~1.9]{Hirche_Key}. For completeness, give a direct proof in SI \ref{sec:proof_one_way_key}. This result might be of independent interest.

It remains to show that the Choi state is a degradable state. We reproduce the argument for completeness: 
A bipartite state $\rho^{AB}$, with purification $\ket{\phi^{ABE}}$, is called degradable if and only if there exists a quantum channel $\mathcal R^{BE}$ that satisfies $\mathcal R^{BE}(\rho^{AB})=\text{tr}_B(\ketbra{\phi^{ABE}})$. Consequently, the Choi state of a degradable channel is a degradable state with the same degrading map. 

It follows that Figures \ref{fig:quantum_capacity} and \ref{fig:quantum_capacity_p} can be interpreted as depicting respectively the one and two-way distillable entanglement (and secret key) of the family of Choi states.

{\bf A sufficient condition for noise resource-contextuality. }
Let us now show that the contextuality of noise is not restricted to the complementary of dephrasure channels in the degradable regime. For showing noise-contextuality it is sufficient that the one-way and two-way capacities have upper and lower bounds with opposite behavior. If this holds, there exists an infinite discrete one-parameter family of channels $\{\cN_{x_n}\}_{n\in\mathbb N}$ for which
$\cQ(\cN_{x_m})<\cQ(\cN_{x_n})
    \text{ and } \cQ_\leftrightarrow(\cN_{x_m})>\cQ_\leftrightarrow(\cN_{x_n})$
for all $n<m$ and decreasing sequence $\{ x_n \}_{n \in \mathbb{N}}$ (so that $x_n > x_m$).  The analogous statement holds for the private capacity.
For the precise statement and proof we refer to SI~\ref{appx:1waycapcont} and Proposition \ref{prop:discretefamily}. In the following we show an example of application of Proposition \ref{prop:discretefamily}. 

We now characterize the behavior of $\cQ,\;\cQ_{\leftrightarrow},\;\cP,\;\cP_{\leftrightarrow}$ of $\cN_{\lambda, \; p}$ in the regime $\lambda>1/2$. Note that, the channel is not degradable for $\lambda>1/2$ and the coherent information of $\cN_{\lambda, \; p}$ is known to be superadditive  \cite{Vikesh_superadditivity}. 

Following the previous discussion, also in this region ($\lambda>1/2$), the two-way assisted quantum and private classical capacity $\cQ_\leftrightarrow(\cN_{\lambda,p})=\cP_\leftrightarrow(\cN_{\lambda, p})=1-\lambda$. 

We bound the one-way capacities from below with the coherent information for one use of the channel as follows (see SI \ref{sec:cohinfoeval}): 
\begin{equation} \label{eq:lb_oneway_cap}
    \cP(\cN_{\lambda, \; p})\geq\cQ(\cN_{\lambda, \; p}) \geq
    \max \{0,1-\lambda(2-H(p)) \}.
\end{equation}
This lower bound can be made tighter as shown in \cite{Vikesh_superadditivity}. However, for our purposes \eqref{eq:lb_oneway_cap} is sufficient.

We obtain an upper bound using a continuity argument. Observe that $\cN_{\lambda,\;p}$ is $\epsilon$-close to the antidegradable channel $\cT(\cdot):=\lambda \Tr(\cdot) \ketbra{\varphi_o} \oplus (1-\lambda) \cI$. The strong continuity property of the one-way quantum capacity~\cite{continuity_capacity}, gives us the following upper bound:
\begin{align*} \label{eq:upper_bound_superadd}
    \cQ(\cN_{\lambda, \; p}) &\leq\cP(\cN_{\lambda, \; p})  \leq \min \{ 1-\lambda, 
     16\lambda \sqrt{p(1-p)} + \\
    &+(4+8\lambda\sqrt{p(1-p)})H( \frac{4\lambda\sqrt{p(1-p)}}{2+4\lambda\sqrt{p(1-p)}}) \}. \numberthis
\end{align*}
With the above bounds, we also show a similar context-dependent behavior of the two capacities when $\lambda>1/2$ (see SI~\ref{appx:1waycapcont} for further details and proofs). In particular, we show that there exists an infinite discrete one-parameter family of channels $\{\cN_{x_n}\}_{n\in\mathbb N}$ for which 
$\cQ(\cN_{x_m})<\cQ(\cN_{x_n}) 
    \text{ and } \cQ_\leftrightarrow(\cN_{x_m})>\cQ_\leftrightarrow(\cN_{x_n})$
for all $n<m$ and decreasing sequence $\{ x_n \}_{n \in \mathbb{N}}$ (so that $x_n > x_m$).  The analogous statement holds for the private capacity.

Finally, for completeness, we characterize the one-way and two-way assisted quantum capacity of the complementary channel $\overline{\cN_{\lambda, \; p}}$ in SI~\ref{app:comp_cap}. 

{\bf An analogy with classical wiretap channels.} The behavior that we have observed for quantum channels is not possible for classical communication over discrete memoryless classical channels. It  is known that feedback does not increase the capacity of a discrete memoryless channel~\cite{shannon1956zero} and, consequently, one- and two-way capacities coincide \cite{shannon1956zero}. 

Nevertheless, we find a meaningful analogy in the context of wiretap channels. Consider a sender, Alice, transmitting information to Bob and Eve through a noisy classical channel. Her task is to send information encoded in such a way that Bob can decode it but Eve learns nothing about the message. The capacity of a wiretap channel $\cN$ is the maximum rate at which the task can be achieved. We consider two different versions; the one- and two-way capacity depending on whether one-way or two-way public classical communication is a free resource which we respectively denote as $\cC(\cN)$ and $\cC_\leftrightarrow(\cN)$.  

\begin{figure}[h]
    \centering
    \includegraphics[width=8.3cm]{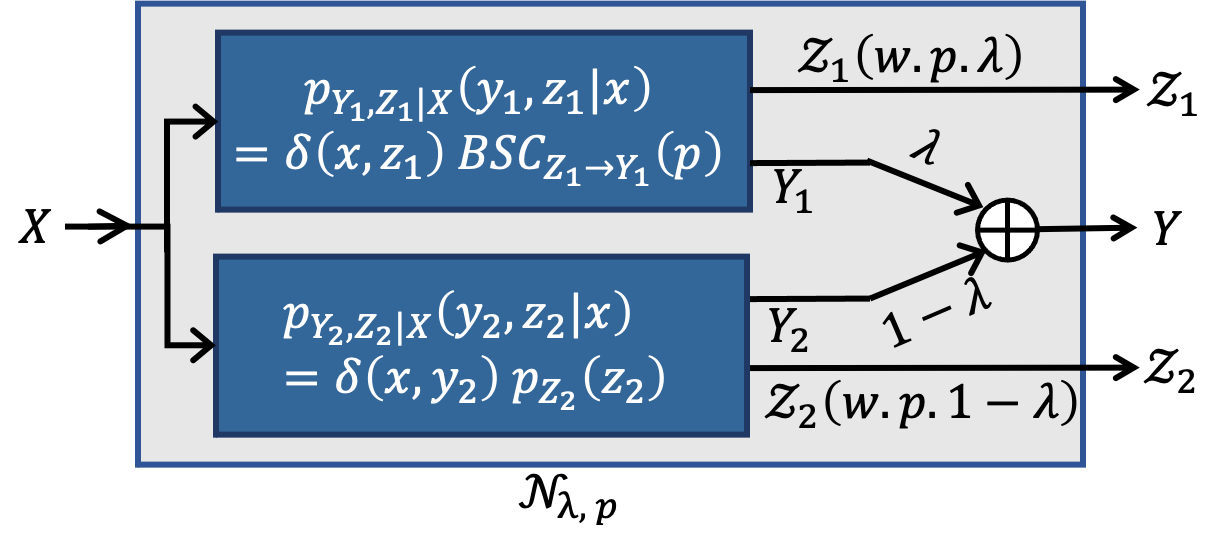}
        \caption{Weighted direct sum classical wiretap channel. With probability $1-\lambda$ the channel gives Bob Alice's input and Eve a random outcome. With probability $\lambda$ it gives Eve Alice's input and Bob a noisy version of Alice's input.}
    \label{fig:classical_channel}
\end{figure}

We exhibit the general model of the channel in Figure~\ref{fig:classical_channel}.
The channel has input $X \in \cX$ and outputs $Y\in \cY$, $Z \in \cZ$  along with a broadcast parameter $L\in \mathcal L$ to both receivers. All the alphabets are binary. 
We associate the receiver obtaining $Y$ with `Bob' and the receiver getting $Z$ with `Eve'. The broadcast parameter takes the value $L=1$ with probability $\lambda$. In this case, the channel behaves like a Markov chain $X \longrightarrow Z_1 \longrightarrow Y_1$  with probability $\lambda$. More precisely, $p_{Y_1,Z_1|X}(y,z|x):= \delta(x,z_1) BSC_{Z_1 \to Y_1}(p)$, where $BSC_{Z_1 \to Y_1}(p)$ represents the binary symmetric channel from $Z_1$ to $Y_1$ with crossover probability $p$. 
With probability $1-\lambda$, $L$ takes the value $2$ and the channel  behaves like $p_{Y_2,Z_2|X}(y,z|x):=\delta(x,y)p_{Z_2}(z)$. 

\begin{figure}[t]
    \centering
    \includegraphics[width=8.3cm]{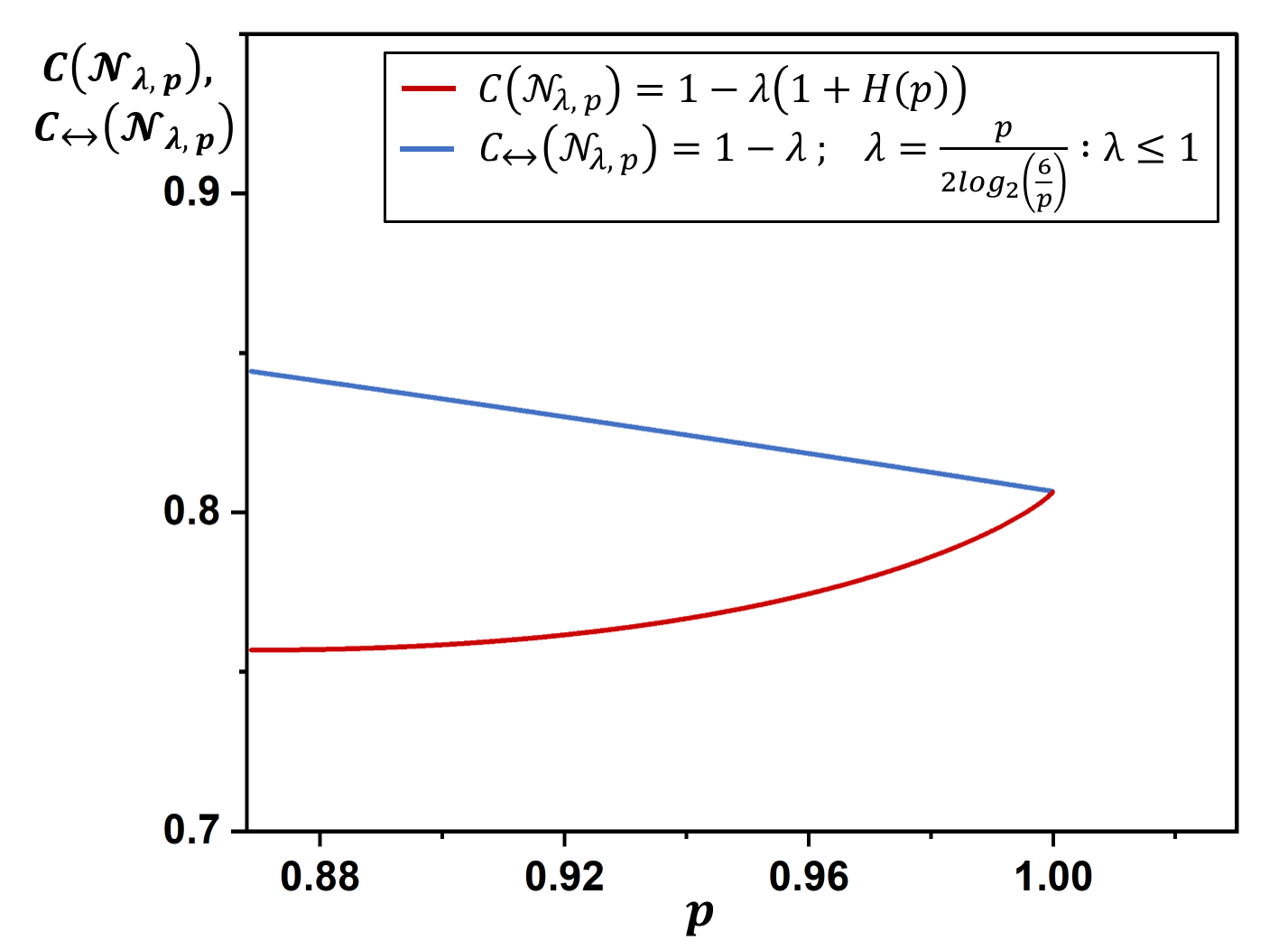}
        \caption{One- vs two-way capacity of the wiretap channels $\cN_{\lambda, \; p}$ as a function of $p$ when  $\lambda(p)=p/(2\log (6/p))$. In the range  $p\in[0.8687,1]$ the one-way capacity monotonically increases while the two-way capacity decreases.}
    \label{fig:classical_capacities}
\end{figure}

We are able to characterize the one- and two-way capacity of the wiretap channel. First, for $\lambda\in[0,1/2]$:
    $\cC(\cN_{\lambda, \; p})=1-\lambda(1+H(p))$, and for $\lambda\in[0,1]$: $\cC_\leftrightarrow(\cN_{\lambda, \; p}) = 1-\lambda$.
We obtain the desired behavior (see Figure~\ref{fig:classical_capacities}), that is the one-way capacity monotonically increases while the two-way decreases, by choosing $\lambda(p)=p/(2\log(6/p))$ and $p\in [0.8687,1]$.

{\bf Proof sketch of the wiretap channel capacities.} Let us start with the one-way capacity. For this, we first observe that the channel $\cN_{\lambda, \; p}$ is degraded~\cite[Proposition~13.2.1]{wilde2013quantum} (which is a classical analogue to a quantum channel being degradable). A channel is degraded if $X$,
$Y$ and $Z$ form the following Markov chain: $X  \longrightarrow Y \longrightarrow Z$. When $\lambda\leq 1/2$, the following stochastic map, allows Bob to produce $Z$ from $Y$. If $L=1$, then Bob provides to Eve $L=2$ and a value $z$ following the probability distribution $p_{Z_2}(z)$. If $L=2$, then with probability $\lambda/(1-\lambda)$ Bob provides to Eve $L=1$ and $z=y$ and with the complementary probability he provides  $L=2$ and $z$ following the probability distribution $p_{Z_2}(z)$.

For degraded channels \cite{Wyner}, the capacity is given by the simple expression:
$\cC(\cN_{\lambda, \; p})=\max_{p_X(x)} \left[ I(X;Y) - I(X;Z) \right].$ 
Let us first expand the difference between the mutual information as a function of the value of the broadcast parameter: $I(X;Y)-I(X;Z) = \lambda \left( I(X;Y_1)-I(X;Z_1) \right)+ (1-\lambda) \left( I(X;Y_2) - I(X;Z_2)\right)$.

We can further expand the expressions: $I(X;Y_1)=  H(X) - H(p) $, $I(X;Z_1)=H(X)$ and $I(X;Y_2)=H(X)$, $I(X;Z_2)=0$. 
Substituting these values we get:
$I(X;Y)-I(X;Z)= (1-\lambda) H(X) - \lambda H(p)$.

The above quantity is maximized when $X$ is uniformly distributed and hence $H(X)=1$. This implies that $\cC(\cN_{\lambda, \; p})=1-\lambda (1+H(p))$ as claimed. 

We now turn to the two-way capacity of the wiretap channel. We first prove that the rate $1-\lambda$ is achievable. When Bob receives $Y=Y_1$, Bob discards it and sends a re-transmission request to Alice. If $Y=Y_2$ then Bob accepts and reports this back to Alice who aborts further transmission so that Eve cannot learn any more. Since, $p_{Y_2|X}(y|x)=\delta(x,y)$, Bob is able to decode $x$ correctly from $Y_2=y$ with probability $1-\lambda$. Note that $Z_2 \indep X$ and since Bob aborts on $Y_1$, Eve cannot learn more than Bob. Thus, the number of private bits that Alice can convey reliably and to Bob per channel use is $1-\lambda$ and hence is an achievable rate.

We now prove the converse, that is, $1-\lambda$ is also an upper bound as follows:
From the channel $X \longrightarrow Z_1 \longrightarrow Y_1$, we see that $Y_1$ is a degraded version of $Z_1$ and hence in this case Bob cannot learn more than Eve, so the optimal strategy for Bob is to ignore it and inform Alice. On the other hand, the second channel ensures that $Y_2=X$ and the decoder can output $Y=Y_2$ to infer $X$ correctly without any further communication from Alice. 
Thus, if Alice does not abort and sends more than $1-\lambda$ bits per channel use then Eve can learn as much as Bob compromising the privacy. Hence the two-way assisted capacity is also upper bounded by $1-\lambda$. Combining the achievability and the converse part, we conclude that $\cC_\leftrightarrow(\cN_{\lambda, \; p}) = 1-\lambda$.

{\it Ackowledgements.-} We thank Kenneth Goodenough for his feedback on the manuscript. We are grateful for the help and support for making graphics for Figure \ref{fig:pop_summary} provided by Pavel Puchenkov from the Scientific Computing and Data Analysis section of Core Facilities at OIST. SS acknowledges support from the Royal Society University Research Fellowship. AN acknowledges support from MEXT Quantum Leap Flagship Program (MEXT QLEAP) Grant No. JPMXS0120319794.

\bibliography{arXiv_final}

\onecolumngrid 

\newpage
\appendix{\Large Supplemental Information}

\section{XZ invariance of the $I_c$ of $\cN_{\lambda, \; p}$}\label{prop:Icoh_invariance}
For completeness, we show that the coherent information of $\cN_{\lambda, \; p}$ is invariant by conjugating the input state with a qubit Pauli $X$ or $Z$.

Let us first start with the action of the channel on a state conjugated by Pauli $Z$:
\begin{align*}
    \cN_{\lambda, \; p}(Z \rho Z) &=\lambda \overline{\cD_p} (Z \rho Z) \bigoplus (1-\lambda) Z\rho Z    \\
    &=\lambda \left[ \bra{0} (Z \rho Z)\ket{0} \varphi_0 + \bra{1} (Z \rho Z)\ket{1} \varphi_1 \right] \bigoplus (1-\lambda) Z \rho Z\\
    &= \lambda \left[ \bra{0} \rho \ket{0} \varphi_0  + \bra{1} \rho \ket{1} \varphi_1  \right] \bigoplus (1-\lambda) Z \rho Z\\
\end{align*}
Thus from Equation~\eqref{eq:entropy_N} we get that
\begin{align*}
    H(\cN_{\lambda, \; p}(Z \rho Z)) &= H(\lambda) + \lambda H(\overline{\cD_p}(\rho)) + (1-\lambda) H(Z \rho Z)\\
    & = H(\lambda) + \lambda H(\overline{\cD_p}(\rho)) + (1-\lambda) H(\rho )\\
    & = H(\cN_{\lambda, \; p}( \rho ))
\end{align*}
Similarly, $\overline{\cN_{\lambda, \; p}}(Z \rho Z) := \lambda \cD_p (Z \rho Z) \bigoplus (1-\lambda) \Tr(Z \rho Z) \ketbra{f} = \overline{\cN_{\lambda, \; p}}(\rho)$, since $\cD_p(Z \rho Z)=\cD_p(\rho)$, as $\cD_p$ is the dephasing channel in Pauli $Z$-bases. Thus, $H(\overline{\cN_{\lambda, \; p}}(Z \rho Z))=H(\overline{\cN_{\lambda, \; p}}(\rho))$ and hence,\\
\begin{align*} \label{eq:Icoh_Zequiv}
I_c(\cN_{\lambda, \; p},Z \rho Z) & = H(\cN_{\lambda, \; p}, Z \rho Z) -  H(\overline{\cN_{\lambda, \; p}}(Z \rho Z))\\ 
& =  H(\cN_{\lambda, \; p}, \rho ) -  H(\overline{\cN_{\lambda, \; p}}( \rho ))\\
&=I_c(\cN_{\lambda, \; p}, \rho) \numberthis
\end{align*}
Let us now consider the action of the channel on a state conjugated by Pauli $X$:
\begin{align*}
    \cN_{\lambda, \; p}(X \rho X) &=\lambda \overline{\cD_p} (X \rho X) \bigoplus (1-\lambda) X \rho X    \\
    &=\lambda \left[ \bra{0} (X \rho X)\ket{0} \varphi_0 + \bra{1} (X \rho X)\ket{1} \varphi_1 \right] \bigoplus (1-\lambda) X \rho X\\
    &\overset{a}{=} \lambda \left[ \bra{1} (\rho)\ket{1} Z \varphi_1 Z + \bra{0} (\rho)\ket{0}  Z \varphi_0 Z  \right] \bigoplus (1-\lambda)  X \rho X  \\
    & = Z \left[ \lambda \left( \bra{1} (\rho)\ket{1} \varphi_0  + \bra{1} (\rho)\ket{1} \varphi_1 \right) \right] Z \bigoplus (1-\lambda)  X \rho X \\
\end{align*}
where (a) holds as $X \ket{0}=\ket{1}, \; X \ket{1}=\ket{0}$, $Z  \ket{\varphi_0}=\ket{\varphi_1}$ and $Z \ket{\varphi_1}=\ket{\varphi_0} $. From Equation~\eqref{eq:entropy_N},
\begin{align*}
    H(\cN_{\lambda, \; p}(X \rho X)) &= H(\lambda) + \lambda H(Z \overline{\cD_p}(\rho) Z) + (1-\lambda) H(X \rho X)\\
    & = H(\lambda) + \lambda H(\overline{\cD_p}(\rho)) + (1-\lambda) H(\rho )\\
    & = H(\cN_{\lambda, \; p}( \rho ))
\end{align*}
Since $X \cD_p(X \rho X) X = (1-p) \rho + p (-X) ZX \rho XZ(-X) = \cD_p(\rho)$ (as $\cD_p$ is the dephasing channel in Pauli $Z$-bases), therefore from Equation~\eqref{eq:entropy_complement_N} we get
$$
H(\overline{\cN_{\lambda, \; p}}(X \rho X))= H(\lambda) + \lambda H(X \cD_p(X \rho X) X) = H(\overline{\cN_{\lambda, \; p}}(\rho)).
$$
Thus,\\
\begin{align*} \label{eq:Icoh_Xequiv}
I_c(\cN_{\lambda, \; p},X \rho X) & = H(\cN_{\lambda, \; p}, X \rho X) -  H(\overline{\cN_{\lambda, \; p}}(X \rho X))\\ 
& =  H(\cN_{\lambda, \; p}, \rho ) -  H(\overline{\cN_{\lambda, \; p}}( \rho ))\\
&=I_c(\cN_{\lambda, \; p}, \rho) \numberthis
\end{align*}

\section{Proof that $\cN_{\lambda, \; p}$ is degradable}\label{prop:degraded_channel} 
\begin{proposition} 
The channel $\cN_{\lambda, \; p}$ given in Equation~\eqref{eq:N(rho)} is degradable for $\lambda \in [0,1/2]$.
\end{proposition}
\begin{proof}
We provide the following proof for completeness. Let $V_{\cN_{\lambda, \; p}}^{A \to BC}$ denote the isometric extension of the channel $\cN_{\lambda, \; p}$. $V_{\cN_{\lambda, \; p}}$ can be written as:
$$
V_{\cN_{\lambda, \; p}}=\sqrt{\lambda} \left\{ \sqrt{1-p} \ket{0}^{B_1} \otimes I^{A \to C} + \sqrt{p} \ket{1}^{B_1} \otimes Z^{A \to C}  \right\} \bigoplus \sqrt{1-\lambda} I^{A \to B_2} \otimes \ket{f}^C
$$
where $|A|=|B_1|=|B_2|=2, B=B_1 \bigoplus B_2$; the corresponding channel is $\cN_{\lambda, \; p}(\rho^A)=\Tr_C \left[ V_{\cN_{\lambda, \; p}} \rho^A V_{\cN_{\lambda, \; p}}^\dagger\right]$, for any input $\rho^A$.
We need to prove that there exists a map $\cR$ such that $\cR \circ \cN_{\lambda, \; p}=\overline{\cN_{\lambda, \; p}}$. We will prove this by showing that $\cR$ is a measure-and-prepare channel. 

For this, we can leverage the direct sum structure of the channel. Since $B=B_1 \bigoplus B_2$, there exists a measurement that distinguishes between the two spaces. The degrading map performs a  different action depending on the measurement outcome. 
    
The measurement outcome will correspond to $B_1$ with probability $\lambda$, the post measurement state is $\bar{D}_p(\rho)$. In this case, the degrading map action consists in preparing the state $\ketbra{f}$. The resulting `substate' is $\lambda \ketbra{f}$;

The measurement outcome will correspond to $B_2$ with probability $1-\lambda$ and the postmeasurement state is $\rho$. In this case, with probability $x \in [0,1]$ the degrading map action consists in preparing the state $\ketbra{f}$ and with probability $1-x$ applying the dephasing map $\cD_p$ to $\rho$. The resultant substate is $(1-\lambda)x \ketbra{f} \bigoplus (1-\lambda)(1-x) \cD_p(\rho)$.

Since $\lambda \in [0,1/2]$ we can choose $(1-\lambda)x + \lambda= 1-\lambda$, which explicitly gives $x=\frac{1-2\lambda}{1-\lambda}$ and also $(1-\lambda)(1-x)=\lambda$.\\
The overall action of the degrading map thus prepares the state:
\begin{eqnarray*}
&  &
(\lambda \ketbra{f} + (1-\lambda)x \ketbra{f}) \bigoplus (1-\lambda)(1-x) \cD_p(\rho)\\
&  &
= (1-\lambda) \ketbra{f} \bigoplus \lambda \cD_p(\rho)  = \overline{\cN_{\lambda, \; p}}(\rho)
\end{eqnarray*}
We thus constructed a measure-and-prepare channel $\cR$ such that $\cR(\cN_{\lambda, \; p}(\rho))= \overline{\cN_{\lambda, \; p}}(\rho)$ for any input $\rho$. A similar proof can also be found in \cite[Appendix~C]{leditzky_dephrasure}.
\end{proof}

\section{Variation of $\cQ(\cN_{\lambda, \; p})$ for $\lambda \leq 1/2$} \label{Appendix:comp_deg}
In this section, we provide intuition of how to find an appropriate parametrization of $p$. Since the two-way capacity $\cQ_{\leftrightarrow}(\cN_{\lambda, \; p})$ for $\lambda \leq 1/2$ (that is when it is additive) is $1-\lambda$ (independent of $p$) we need to show that $\cQ(\cN_{\lambda, \; p})$ increases as a function of $\lambda$ with $p$ being a function of $\lambda$, in the range $p(\lambda) \in (0,1/2)$. 

To see where $\cQ(\cN_{\lambda, \; p})$ increases, we use elementary differentiation for Equation~\eqref{eq:deg_cap} by treating the parameter $p$ as a function of $\lambda$, denoted by $p(\lambda)$ as follows:
\begin{eqnarray} 
    \frac{\mathrm{d}}{\mathrm{d}\lambda}\cQ(\cN_{\lambda, \; p}) 
    &  =  &
    \frac{\mathrm{d}}{\mathrm{d}\lambda} \left[ 1-\lambda (2-H(p(\lambda)))\right]\nonumber\\
    &  =  & 
    H(p(\lambda))-2+\lambda \frac{\mathrm{d}}{\mathrm{d}\lambda}p(\lambda) \log \left( \frac{1-p(\lambda)}{p(\lambda)} \right)\nonumber\\
    &\Rightarrow & H(p(\lambda))+\lambda p'(\lambda) \log \left( \frac{1-p(\lambda)}{p(\lambda)} \right) 
      \geq 2. \label{eq:deg}
\end{eqnarray}
 
Since $H(p(\lambda)) \geq 0$:
\begin{eqnarray*}
p'(\lambda) 
& \geq &
\frac{2}{\lambda} \times \frac{1}{\log \left( 1/p(\lambda) - 1\right)}\\
& \geq &
\frac{2}{\lambda} \times \frac{1}{\log \left( 1/p(\lambda) + 1\right)}\\
& \geq &
\frac{2}{\lambda} \times \frac{1}{1/p(\lambda) }\\
& = &
2\frac{p(\lambda)}{\lambda}
\end{eqnarray*}
In particular, the choice of $p(\lambda)=4\lambda-1$, for $\lambda \in [1/4,3/8)$, fulfills the condition of the above inequality. The behavior for this choice is depicted in Figure~\ref{fig:quantum_capacity}.

\section{An upper bound for $ E_{r}(\overline{\cN_{\lambda, \; p}})$ }\label{sec:entropyupperbound}
For completeness, we reproduce the definition and selected properties of the relative entropy of entanglement. 
\begin{definition}\label{def:entanglement_entropy}
Let $\textrm{SEP}$ denote set of all separable bipartite states. The relative entropy of entanglement of a bipartite state $\rho^{AB}$ is defined as:
\[
E_r(\rho^{AB}):=\inf \limits_{\sigma^{AB}: \sigma^{AB} \in \text{ SEP }} D(\rho^{AB}||\sigma^{AB})
\]
Similarly, the relative entropy of entanglement of a channel $\cN_{\lambda,\;p}$ is defined as:
\[
E_r(\cN_{\lambda, \; p}):=\sup \limits_{\ket{\phi}^{AA'}:\Tr_{A'}(\phi^{AA'})=\rho^A} E_r [(\cI^A \otimes \cN_{\lambda, \; p}^{A' \to B})(\phi^{AA'})]
\]
\end{definition}

\begin{fact} \mbox{\cite[Proposition~3]{E_rconvexity}}\label{fact:convexity_Er}
The relative entropy of entanglement is convex. For all $a \in [0, 1]$ and all density operators $\rho_1,\; \rho_2$ it holds that
\[
E_r(a\rho_1 + (1 - a)\rho_2) \leq aE_r(\rho_1) + (1 - a)E_r(\rho_2).
\]
\end{fact}

We upper bound relative entropy of entanglement of $\overline{\cN_{\lambda, \; p}}$ as follows:
\begin{align*}
    E_{r}(\overline{\cN_{\lambda, \; p}}) & \overset{a}{=} E_{r} (\rho^{AB}_{\overline{\cN_{\lambda, \; p}}})\\
    & \overset{b}{=}\inf\limits_{\sigma^{AB} \in SEP} \left[ D\left( \left\{ \lambda (\cI^A \otimes \cD_p)(\phi_{\overline{\cN_{\lambda, \; p}}}^{AA'}) \bigoplus (1-\lambda) (\cI^A \otimes \Tr)(\phi^{AA'}_{\overline{\cN_{\lambda, \; p}}})(I \otimes \ketbra{f})\right\}\Bigg\Vert\sigma^{AB} \right) \right]\\
    & \overset{c}{=} \inf\limits_{\sigma^{AB} \in SEP} \left[ D\left( \left\{ \lambda \rho_1^{AB} \bigoplus (1-\lambda) (\rho^A \otimes \ketbra{f}^B)\right\} \Bigg\Vert \sigma^{AB} \right) \right]\\
    &= \inf\limits_{\sigma^{AB} \in SEP} \left[ D\left( \left\{ \lambda \left(\rho_1^{AB} \bigoplus \ketbra{0}^{AB} \right) + (1-\lambda) \left(\ketbra{0}^{AB} \bigoplus \rho^A \otimes \ketbra{f}^B\right)\right\} \Bigg\Vert \sigma^{AB} \right) \right]\\
    &= E_r\left[ \lambda \left(\rho_1^{AB} \bigoplus \ketbra{0}^{AB} \right) + (1-\lambda) \left(\ketbra{0}^{AB} \bigoplus \rho^A \otimes \ketbra{f}^B\right) \right]\\
    & \overset{d}{\leq} \lambda E_r(\rho_1^{AB})+(1-\lambda)E_r(\rho^A \otimes \ketbra{f}^B)\\
    & \overset{e}{\leq} \lambda E_r(\cD_p)\\
    &\overset{f}{=} \lambda I_c(\cD_p)=\lambda(1-H(p))
\end{align*}
where (a) and (b) follows by choosing $\phi_{\overline{\cN_{\lambda, \; p}}}$ to be the purification of an input that achieves the maximum for $E_r(\overline{\cN_{\lambda, \; p}})$ in Definition~\ref{def:entanglement_entropy} and $\rho^{AB}_{\overline{\cN_{\lambda, \; p}}}:=(\cI^A\otimes \overline{\cN_{\lambda, \; p}}^{A'\to B})(\Phi_{\overline{\cN_{\lambda, \; p}}})^{AA'}$;\\
(c) follows from the following identification: \\
\vspace{-0.3cm}
$$
\rho_1:= (\cI^A \otimes \cD_p^{A' \to B}(\phi_{\overline{\cN_{\lambda, \; p}}}^{AA'})) \text{ and } \rho:=\Tr_{A'}(\phi_{\overline{\cN_{\lambda, \; p}}}^{AA'});
$$
(d) follows by the convexity of the relative entropy~\cite[Proposition~3]{E_rconvexity} ;\\
(e) follows since $E_r(\rho^A \otimes \ketbra{f}^B)=0$ and $\rho_{\overline{\cN_{\lambda, \; p}}}$ might not be the maximizing state for $E_r(\cD_p)$.\\
(f) follows from \cite[Equation~39]{entanglemententropy_capacity}.

\section{Proof of the one-way capacity of $\cN_{\lambda, \; p}$ for $\lambda\in[0,1/2]$ and two-way capacity of $\cN_{\lambda, \; p}$ for $\lambda\in[0,1]$}\label{proof_theorem1}
We prove the one-way capacity in Lemma \ref{lem:one_way_quantum} and two-way capacity in Lemma \ref{lem:two_way_converse}.

\begin{lemma}
\label{lem:one_way_quantum}
The one-way classically assisted quantum and private capacities of $\cN_{\lambda, \; p}$ when $\lambda \in [0,1/2]$ are given by 
$$\cP(\cN_{\lambda, \; p})=\cQ(\cN_{\lambda, \; p})=1-\lambda(2-H(p)).
$$
\end{lemma}
\begin{proof}
Given that $\cN_{\lambda, \; p}$ is degradable (see SI~\ref{prop:degraded_channel} for the proof), its quantum and private capacities coincide \cite{Degradable_capacity}. Moreover, the quantum (and private) capacity is given by the coherent information (by \cite[Theorem~2]{Degradable_capacity}, also \cite{Devetak_dephasing}).
We now evaluate the coherent information of $\cN_{\lambda, \; p}$ and then maximize it over all the input density operators to get the capacity. Using the definition of $\cN_{\lambda, \; p}$ and $\overline{\cN_{\lambda, \; p}}$ we can rewrite:
\begin{align*} \label{eq:coherentinfo}
    I_c(\cN_{\lambda, \; p}) &= \max \limits_{\rho^A} \left[ H(\cN_{\lambda, \; p}(\rho))- H(\overline{\cN_{\lambda, \; p}}(\rho)) \right] \numberthis
\end{align*}
Since, $\cN_{\lambda, \; p}$ is a combination of the complementary dephasing channel and identity channel, for any input $\rho$ the following equalities hold (proved in SI \ref{prop:Icoh_invariance}):
\begin{align} \label{eq:equal_Ic}
    I_c(\cN_{\lambda, \; p}, \rho) = I_c(\cN_{\lambda, \; p}, Z\rho Z) = I_c(\cN_{\lambda, \; p}, X \rho X).
\end{align}
The coherent information of a degradable channel is concave with respect to input density operators \cite[Theorem~13.5.2]{wilde2013quantum} and using Equation~\eqref{eq:equal_Ic}, we get:
\begin{align*}
    I_c(\cN_{\lambda, \; p}, \rho) &= \frac{1}{2}I_c(\cN_{\lambda, \; p}, \rho) +\frac{1}{2}I_c(\cN_{\lambda, \; p}, Z \rho Z) \leq I_c(\cN_{\lambda, \; p}, \frac{\rho + Z \rho Z}{2}). 
\end{align*}
The above equation shows that $I_c(\cN_{\lambda, \; p}, \rho )$ is upper bounded by the coherent information of $\cN_{\lambda, \; p}$ evaluated over the states diagonal in the $Z$ basis, such as $\frac{\rho + Z \rho Z}{2}$. Hence, $I_c(\cN_{\lambda, \; p})$ is maximized for an input $\rho^A$ that is diagonal in the $Z$ basis.
Similarly,
\begin{align*}
    I_c(\cN_{\lambda, \; p}, \rho) &= \frac{1}{2}I_c(\cN_{\lambda, \; p}, \rho) +\frac{1}{2}I_c(\cN_{\lambda, \; p}, X \rho X ) \leq I_c(\cN_{\lambda, \; p}, \frac{\rho + X \rho X}{2}) 
\end{align*}
shows that $I_c(\cN_{\lambda, \; p}, \rho )$ is maximized for a $\rho^A$ which is diagonal in the $X$ basis.
Hence, we get that the maximum in $I_c(\cN_{\lambda, \; p})$ is attained for an input density operator that is  simultaneously diagonal in both $X$ and $Z$-bases. This implies that maximum is achieved for $\rho^A = \pi^A$.\\

Substituting Equations~\eqref{eq:entropy_N} and \eqref{eq:complement_N} in Equation~\eqref{eq:coherentinfo} we get the desired result:
\begin{align*} \label{eq:deg_cap}
    I_c(\cN_{\lambda, \; p}, \pi) &= (1-\lambda) H(\pi) +\lambda \left[ H(\overline{\cD_p}(\pi))  -  H(\cD_p(\pi)) \right]\\
    & \overset{a}{=} (1-\lambda) - \lambda (1-H(p)) = 1-\lambda(2-H(p)) \numberthis
\end{align*}
where (a) follows due to:
\begin{itemize}
    \item $ H(\overline{\cD_p}(\pi))=H(\mathrm{diag}[1-p,\;p]))=H(p)$;
    \item $ H(\cD_p(\pi))=H(\pi)=1 $.
\end{itemize}
\end{proof}

\begin{lemma} \label{lem:two_way_converse}
For all $\lambda \in [0,1]$: ${\cQ}_\leftrightarrow(\cN_{\lambda, \; p})={\cP}_\leftrightarrow(\cN_{\lambda, \; p})=1-\lambda$.
\end{lemma}
\begin{proof}
    Since $\cN_{\lambda, \; p}$ is teleportation stretchable, by \cite[Theorem~5]{entanglemententropy_capacity}, the two-way classically assisted private capacity and hence the quantum capacity is upper bounded by the relative entropy of entanglement of $\cN_{\lambda, \; p}$ (formally defined in Section~\ref{sec:entropyupperbound} of the SI).
    We thus have,
    \begin{equation} \label{eq:two_way_ubE_r}
        {\cQ}_\leftrightarrow(\cN_{\lambda, \; p}) \leq E_r(\cN_{\lambda, \; p})
    \end{equation}
    where
    \begin{align*}
       E_r(\cN_{\lambda, \; p}) &:= \sup\limits_{\Phi^{AA'}: \text{pure}} E_r \left[ \left\{ \cI \otimes \cN_{\lambda, \; p} \right\} (\Phi) \right] \\
       &=\sup\limits_{\Phi^{AA'}: \text{pure}} E_r[ \lambda \left\{ \cI \otimes \overline{\cD_p} \right\} (\Phi)  \bigoplus (1-\lambda) \left\{ \cI \otimes \cI \right\} (\Phi) ]\\
       & \overset{(a)}{\leq} \lambda \sup\limits_{\Phi^{AA'}: \text{pure}} E_r\left[ \left\{ \cI \otimes \overline{\cD_p} \right\} (\Phi) \right] + (1-\lambda) \sup\limits_{\Phi^{AA'}: \text{pure}} E_r\left[ \lambda \left\{ \cI \otimes \cI \right\} (\Phi) \right]\overset{(b)}{=} 1-\lambda
    \end{align*}
    where, (a) holds by applying convexity of the relative entropy~\cite[Proposition~3]{E_rconvexity} and (b) holds by the observation that $\overline{\cD_p}(\rho)$ is separable for all $\rho$ so $E_r(\overline{\cD_p})=0$ and $E_r(\cI)=1$ by \cite[Proposition~1]{rel_ent_entropy_calc}.
    Combining above with Equation~\eqref{eq:two_way_ubE_r} gives the desired upper bound.

    The upper bound is achievable. This can be observed by inspection; the encoder can send half of a maximally entangled state. This procedure will prepare a joint maximally entangled state when the channel acts as the identity which occurs with probability $1-\lambda$. Exploiting the direct-sum structure of the channel, the communicating parties can distinguish between the action of $\overline{\cD_p}$ and $I$ and consume the maximally entangled states to communicate noiselessly at a rate of $1-\lambda$. 
\end{proof}

\section{Characterizing the one-way capacity using continuity when $\lambda \in [1/2,1]$}\label{appx:1waycapcont}
The one-way quantum and private capacities of the channel from \eqref{eq:N(rho)} are not characterized by the coherent information in this range of $\lambda,p$ \cite{Vikesh_superadditivity}. For this reason, we resort to analyzing the lower and upper bounds for the quantum capacity.
\subsection{Lower bound on $\cQ(\cN_{\lambda, \; p})$ for $\lambda \in [1/2,1]$}\label{sec:cohinfoeval} 
One can consider the coherent information $I_c(\cN, \rho)$ for one use of the channel as a trivial lower bound,
\begin{align}
I_c(\cN, \rho) &= H(\cN_{\lambda, \; p}(\rho)) - H(\overline{\cN_{\lambda, \; p}}(\rho)) \nonumber \\
&=\lambda (H(\overline{\cD_p}(\rho)) - H(\cD_p(\rho)) ) + (1-\lambda) H(\rho) 
\end{align}
For $\rho = \pi$, this turns out to be $1-\lambda(2-H(p))$.
 
\subsection{Upper bound on $\cQ(\cN_{\lambda, \; p})$ for $\lambda \in [1/2,1]$}\label{Appendix:upper_bound}
We start by stating the following continuity property of the one-way classically assisted capacity of almost antidegradable channels:
\begin{definition} \cite{continuity_capacity}
\label{def:appx_antidegrad}
    A quantum channel $\chi^{A \to B}$ is said to be $\epsilon$-close anti-degradable if there exists an anti-degradable channel $\eta^{A \to B}$ such that $\lVert \chi - \eta \rVert_\diamond \leq \epsilon$ where $\lVert \chi - \eta \rVert_\diamond \equiv \text{sup}_n \lVert \chi \otimes \mathbb{I}_n - \eta \otimes \mathbb{I}_n\rVert_{\text{tr}}$ and $\lVert \chi - \eta \rVert_{\text{tr}}\equiv \text{max}_{\rho} \lVert \chi (\rho) - \eta (\rho)\rVert_{\text{tr}}$ with $\lVert . \rVert_{\text{tr}}$ denoting the usual trace norm of an operator.
\end{definition}
\begin{fact}\mbox{\cite[Corollary~A.4]{continuity_capacity}}
    \label{fact:continuity_capacity}
    For a quantum channel $\chi^{A \to B}$ that is $\epsilon$-close anti-degradable it holds that,
    \[
    \cQ(\chi) \leq \cP(\chi) \leq 2 \epsilon \log |B| + 2(2 + \epsilon)H\left(\frac{\epsilon}{2+\epsilon}\right)
    \]
\end{fact}

Building on Definition \ref{def:appx_antidegrad} and Fact \ref{fact:continuity_capacity}, we prove the upper bound for $\cQ(\cN_{\lambda, \; p})$ given in Equation~\eqref{eq:upper_bound_superadd}. 
\begin{proof}[Proof of Equation~\eqref{eq:upper_bound_superadd}] \label{proof:eq6}

We consider the channel $\cT^{A \to B}:= \lambda \Tr(\cdot) \ket{\varphi_0}\bra{\varphi_0} \bigoplus (1-\lambda)\cI$ and show that it is close to the channel $\cN_{\lambda, \; p}$ in the diamond norm. 
For this, we expand any density matrix $\rho^{AR}$ in the computational basis for $A$ and any canonical basis for $R$, as 
$\rho^{AR}=\mathop{\Sigma}_{i,j \in A} \mathop{\Sigma}_{k,l \in R} \rho_{ijkl} \ket{i}\bra{j}^A \otimes \ket{r_k}\bra{r_l}^R$. The following chain of inequalities holds:
\begin{align*}
    \Vert \cN_{\lambda, \; p} - \cT \rVert_\diamond &= \max_{\rho^{AR}} \lVert \left(\cN_{\lambda, \; p}^{A \to B} \otimes \cI^R \right) \rho^{AR} - \left(\cT^{A \to B} \otimes \cI^R \right) \rho^{AR}\rVert_1\\
    &=\max_{\rho^{AR}} \lambda \lVert \Sigma_{{i,j \in A}}\Sigma_{{k,l\in R}} \rho_{ijkl} \left\{ \left(\bra{0}\ket{i}\bra{j}\ket{0} \varphi_0 + \bra{1}\ket{i}\bra{j}\ket{1} \varphi_1\right) -\Tr(\ket{i}\bra{j}) \varphi_0\right\} \otimes \ket{r_k}\bra{r_l}\rVert_1\\
    &\overset{a}{=}\max_{\rho^{AR}} \lambda \lVert \Sigma_{k,l\in R} \rho_{00kl} \varphi_0 \otimes \ket{r_k}\bra{r_l} + \Sigma_{k,l\in R} \rho_{11kl} \varphi_1 \otimes \ket{r_k}\bra{r_l}-\varphi_0\otimes \rho^R \rVert_1\\
    &=\max_{\rho^{AR}} \lambda \lVert  \varphi_0 \otimes\rho^R-\Sigma_{k,l\in R}\rho_{11kl} \varphi_0 \otimes \ket{r_k}\bra{r_l} + \Sigma_{k,l\in R} \rho_{11kl} \varphi_1 \otimes \ket{r_k}\bra{r_l}-\varphi_0\otimes \rho^R \rVert_1\\
    &= \max_{\rho^{AR}} \lambda \lVert  \Sigma_{k,l\in R}\rho_{11kl} (\varphi_1-\varphi_0) \otimes \ket{r_k}\bra{r_l} \rVert_1 \\
    & = \max_{\rho^{AR}} \lambda \lVert  \Sigma_{k,l\in R}\rho_{11kl}\rVert_1 \lVert \varphi_1-\varphi_0 \rVert_1 \\
    &\overset{b}{\leq} \lambda \lVert \varphi_1-\varphi_0 \rVert_1 \\
    &=\lambda 2\sqrt{1-|\bra{\varphi_0}\ket{\varphi_1}|^2} \\
    &\overset{c}{=}4\lambda \sqrt{p(1-p)}   
\end{align*}
where (a) holds by the observation that $\rho^R=\Sigma_{i \in A} \Sigma_{k,l \in R}\rho_{iikl} \ket{r_k}\bra{r_l}^R$;\\
(b) holds as $1=\lVert \rho^R\Vert_1=\lVert \Sigma_{i \in A} \Sigma_{k,l \in R} \rho_{iikl} \ket{r_k}\bra{r_l}\rVert_1 \geq \lVert \Sigma_{k,l \in R} \rho_{11kl} \ket{r_k}\bra{r_l}\rVert_1 $; \\
and (c) holds by direct calculation of the eigenvalues of the operator $\varphi_1-\varphi_0$ and the fact that $\varphi_0=(\sqrt{1-p},\sqrt{p})^T$ and $\varphi_1=(\sqrt{1-p},-\sqrt{p})^T$.
   
It is easy to see that the channel $\cT$ is an antidegradable channel (can be proved by the help of a simple measure-and-prepare map).
   
Finally, by Definition~\ref{def:appx_antidegrad}, the channel $\cN_{\lambda, \; p}$ is $4\lambda \sqrt{p(1-p)}$-close anti-degradable and hence by Fact~\ref{fact:continuity_capacity}, the private and quantum capacities are bounded from above by:
\begin{equation*} 
       \cQ_{ub}(\cN_{\lambda, \; p}):=16\lambda\sqrt{p(1-p)}+2\left(2+4\lambda \sqrt{p(1-p)}\right) H\left( \frac{4\lambda \sqrt{p(1-p)}}{2+4\lambda \sqrt{p(1-p)}} \right).
\end{equation*}
 \end{proof}
\subsection{A discrete family of channels for which the one-way capacity increases and the two-way capacity decreases with $\lambda \in [1/2,1]$}
Let us first show a general statement regarding  how to construct a family of channels for which the one-way and two-way assisted capacities have opposite behavior. For this purpose it is enough to have monotonic bounds on the capacity: 
\begin{proposition} 
Let $a,b\in\mathbb R$ and let $\{\cN_x\}_{x\in[a,b]}$ be a family of quantum channels. 
Let $\cQ_\leftrightarrow(\cN_x)$ be a continuous and strictly monotonically decreasing function of $x \in [a,b]$. Further suppose that there exists an upper and a lower bound on the (one-way) capacity of the form $\cQ_{lb}(\cN_x)\leq\cQ(\cN_x)\leq\cQ_{ub}(\cN_x)<\cQ_\leftrightarrow(\cN_x)$ that are continuous and increase monotonically in $x \in [a,b]$ and that $\cQ_{lb}(\cN_a)=\cQ_{ub}(\cN_a)$. Then there exists a strictly decreasing infinite sequence of values $\{x_n\}_{n\in\mathbb N}$ with $b>x_i>a$ for all $i$ and such that for all $n<m$:
\begin{align}
    &\cQ(\cN_{x_m})<\cQ(\cN_{x_n}) \nonumber \\
    \text{and}~~& \cQ_\leftrightarrow(\cN_{x_m})>\cQ_\leftrightarrow(\cN_{x_n})\ . \nonumber
\end{align}
\label{prop:discretefamily}
    \end{proposition}
    
  \begin{proof}
Let $x_0=b$, we define $x_n$ from $x_{n-1}$ recursively for $n\in\mathbb N$ as follows:
We let the auxiliary variable $t$ take the value such that $\cQ_{ub}(\cN_t)=\cQ_{lb}(\cN_{x_{n-1}})$. The existence of $t<x_{n-1}$ is guaranteed because $\cQ_{lb}(\cN_x),\cQ_{ub}(\cN_x)$ are monotonically increasing and $\cQ_{lb}(\cN_a)=\cQ_{ub}(\cN_a)$. Then $x_n=(t+a)/2$.

Let us now show that the infinite sequence of channel indexed by $n$ verifies the desired properties. 
First, since $x_n$ is monotonically decreasing with $n$, $\cQ_\leftrightarrow(\cN_{x_n})$ is monotonically increasing. 
Second, $\cQ(\cN_{x_n})\geq\cQ_{lb}(\cN_{x_n})>\cQ_{ub}(\cN_{x_{n+1}})\geq\cQ(\cN_{x_{n+1}})$ which completes the proof.
\end{proof}
{\it Observation:} The one-parameter family of channels $\cN_{\lambda(p), p}$ with $\lambda(p) = \frac{1}{2}+p$ in the range $p\in [0,0.0002]$ verifies the conditions of Proposition \ref{prop:discretefamily}. 

First, let us now verify the conditions for the one-way capacity. Let us first analyze the lower bound, which was given in Eq. \eqref{eq:deg_cap}. It particularizes to $1-(1/2+p)(2-H(p))=H(p)/2-2p+pH(p)$. To show that the lower bound is increasing with $p$, let us take the derivative with respect to $p$. The derivative turns out to be
\begin{equation*}
    \frac{1}{2} (-4 - (1 - 4 p) \log[1 - p] - (1 + 4 p) \log[p]).
\end{equation*}
For $p<1/4$, the derivative is larger than $\frac{1}{2} (-4 - \log[p])$, which in turn is larger than zero if $p<1/16$.

The upper bound on the one-way assisted capacity is given by \eqref{eq:upper_bound_superadd}, which for the given parametrization reduces to:
\begin{equation}
\cQ_{ub}(\cN_{\lambda(p), \; p}):=(8+16p)\sqrt{p(1-p)}+(4+(4+8p) \sqrt{p(1-p)}) H\left( \frac{(1+2p) \sqrt{p(1-p)}}{1+(1+2p) \sqrt{p(1-p)}} \right).
\end{equation}
The upper bound is monotonically increasing in the range $p\in [0,1/16]$. This follows from the observation that the functions $\sqrt{p(1-p)}$, $8+16p$ as well as $H\left(\frac{(1+2p) \sqrt{p(1-p)}}{1+(1+2p) \sqrt{p(1-p)}}\right)$ are all monotonically increasing with respect to $p$ in this range. 

Second, the two-way capacity is given by  $\cQ_\leftrightarrow(\cN_{\lambda(p), p})=1-\lambda=1/2-p$ which is monotonically decreasing with $p$.  The upper bound on the one-way capacity is smaller than the two-way capacity for $p<0.0002$. 
We end by observing that $\cQ_{ub}(\cN_{\lambda(p), \; p}) = \cQ_{lb}(\cN_{\lambda(p), \; p}) = 0$ at $p=0$.
Therefore, from the above proposition, there exists a discrete infinite family of channels for which the one-way capacity increases and the two-way capacity decreases. 

\section{One-way and two-way assisted capacity of the complementary channel} \label{app:comp_cap}
In this section, we fully characterize the capacity of the channel given by Eq. \eqref{eq:complement_N} in the range $\lambda \in [0,1/2]$; i.e. the complementary channel to our main construction. 

Since $\cN_{\lambda, \; p}$ is degradable for $\lambda \in [0,1/2]$, therefore the complementary channel $\overline{\cN_{\lambda, \; p}}:=\lambda \cD_p \bigoplus (1-\lambda) \ketbra{f}$ is anti degradable and the one-way assisted capacity of ${\cQ}(\overline{\cN_{\lambda, \; p}})=0$ \cite{bennett1997capacities}.

\begin{lemma} \label{lem:2-way_complemantary}
The two-way classically assisted capacity of ${\cal Q}_\leftrightarrow(\overline{\cN_{\lambda, \; p}})=\lambda (1-H(p))$.
\end{lemma}
\begin{proof}
For any input state $\rho^A$, the channel output can be written as:
\[
\sigma^B := \overline{\cN_{\lambda, \; p}(\rho)}:=(1-\lambda) \ketbra{f} \bigoplus \lambda D_p(\rho)
\]
Since the channel is a direct sum channel, let the output space of the `flag' channel (that maps any input to a fixed pure state referred to as the flag) be $B_2$ and the output space of $\cD_p$ be $B_1$, such that $B= B_1 \bigoplus B_2$ and $B_1 \cap B_2 = \{ 0 \}$.\\
\textbf{Achievability: }
When Bob receives the state $\sigma^B$, Bob applies the measurement of form $\{M^{B_1},I-M^{B_1}\}$. If Bob gets a non-zero outcome, he sends an acknowledgement classically to Alice, to abort the protocol. If Bob gets a zero outcome, he knows that he received the outcome of dephasing channel $\cD_p$ and indicates this back to Alice and declares that as the output. This ensures that a rate of $\lambda I_c(\cD_p)$ (which is the probability of getting the outcome from a dephasing channel times its capacity) is achievable.\\
\textbf{Converse: }Since $\overline{\cN_{\lambda, \; p}}$ is teleportation stretchable  \cite{entanglemententropy_capacity}, the relative entropy of entanglement of $\overline{\cN_{\lambda, \; p}}$ upper bounds the two-way assisted private and quantum capacity.

Finally, in SI~\ref{sec:entropyupperbound} we show that $E_{r}(\overline{\cN_{\lambda, \; p}})= \lambda I_c(\cD_p)=\lambda(1-H(p))$.
This proves the claim.
\end{proof}

\section{One-way distillable entanglement and secret key of degradable states}
\label{sec:proof_one_way_key}
The one-way distillable entanglement (respectively secret key) of a bipartite quantum state $\rho^{AB}$, denoted by $D_\rightarrow(\rho^{AB})$ (respectively $K_\rightarrow(\rho^{AB})$) is defined as the maximal rate of distilling entangled bits (respectively secret classical bits) given access to identical and independent copies of $\rho^{AB}$ using local operations and forward (or one-way) classical communication(LOCC) from sender to receiver. We refer to \cite{Devetak_Winter} for the formal definitions.

In general, a one-way LOCC protocol can  be modeled as a quantum instrument $T:A \to A'M$, and is defined as 
\[
\cT^{A \to A'M}(\eta^A):= \sum_m \cT_m^{A \to A'} (\eta^A) \otimes \ketbra{m}^M\,,
\]
where $\ket{m}$ forms an orthonormal basis for the classical register $M$, to be transmitted using forward classical communication and for each $m$ the map $\cT_m: A \to A'$ is completely positive map. 

The closed form expression for $D_\rightarrow(\rho^{AB})$ is given by the following regularized formula \cite{Devetak_Winter}:
\begin{equation} \label{eq:distil_entanglement}
D_\rightarrow(\rho^{AB})= \lim_{n \to \infty} \frac{1}{n} \max_{\cT^{A \to A'M}}I_c(A' > BM)_{\left( (\cT^{A \to A'M} \otimes \cI^B)(\rho^{AB}) \right)}
\end{equation}
\noindent where the maximum is taken over all quantum instruments.

The following fact gives a characterization of the $D_\rightarrow(\rho^{AB})$ for degradable bipartite quantum states.
\begin{fact}\cite[Proposition~2.4]{leditzky2017useful} \label{fact:degradable_distil_ebit}
Given a degradable state $\rho^{AB}$, the one-way distillable entanglement is additive and is given by its coherent information, that is,
\[
D_\rightarrow(\rho^{\otimes n}_{AB})=nD_\rightarrow(\rho^{AB})=nI_c(\rho).
\]
\end{fact}
 We now show that the one-way distillable secret key of a bipartite state is also equal to its coherent information. Recall:
\begin{fact}\cite[Theorem 8]{Devetak_Winter} \label{fact:key}
For any state $\rho^{ABE}$, 
$$
K_\rightarrow(\rho) = \lim_{n \to \infty} K^{(1)}(\rho^{\otimes n}), \textrm{ with } K^{(1)}(\rho) := \max_{\Lambda,T|X} \left[ I(X;B|T)-I(X;E|T) \right], 
$$
where the maximization is over all POVMs $\Lambda = \{\Lambda_x\}_{x \in \mathcal{X}}$ and channels $R$ such that $T = R(X)$. The information quantities are evaluated with respect to the state $\omega^{TXBE} = \Sigma_{t,x} R(t|x)p(x) \ket{t}\bra{t}^T \otimes \ket{x}\bra{x}^X \otimes \Tr_A\left[\rho^{ABE}(\Lambda_x \otimes I^{BE})\right]$. 
\end{fact}
Our observation can be formalised as follows:
\begin{lemma} \label{lem:key}
\label{proof_one_way_key}
For any degradable state $\rho^{AB}$, with the purification $\ket{\rho}^{ABE}$ and the system $E$ holds the purifying register of $\rho^{AB}$, the one-way distillable secret key is given by the coherent information of $\rho^{AB}$, that is,
$$
K_\rightarrow(\rho)=I_{c}(\rho):=H(B)_\rho-H(AB)_\rho
$$
\begin{proof}
    From \cite[Corollary~2]{Horodecki_key}: $K_\rightarrow(\rho) \geq D_\rightarrow(\rho)$, where $D_\rightarrow(\rho)=I_c(\rho)$ (from \cite[Proposition~2.4]{leditzky2017useful}). Thus $K_\rightarrow(\rho) \geq I_c(\rho)$.

    It remains to show that $K_\rightarrow(\rho) \leq I_c(\rho)$. For this, fix any POVM $\Lambda$ and the classical $R$ in Fact~\ref{fact:key}. We now define the following classical-quantum state obtained from the measurement $\Lambda$:
    \[
    \omega^{XBE}:=\sum_x p(x) \ketbra{x}^X \otimes \rho_x^{BE}
    \]
    where $p(x):= \Tr \left[ (\Lambda_x^A \otimes I^{BE}) \ketbra{\rho}^{ABE}\right]$ and $\rho_x^{BE}:=\Tr_A \left[ (\Lambda_x^A \otimes I^{BE}) \ketbra{\rho}^{ABE} \right]$. Now the channel $T$ in Fact~\ref{fact:key} is a classical channel that maps the measurement outcome $x$ to $t$ with probability $R(t|x)$ and therefore the overall state as:
    \begin{align} \label{eq:omega_BEXT}
        \omega^{TXBE}:=\sum_{t,x} R(t|x)p(x) \ketbra{t}^T \otimes \ketbra{x}^X \otimes \rho_x^{BE}\;.
    \end{align}
    We also consider the spectral decomposition $\rho_x^{BE}=\sum_y p(y|x) \ket{\phi_y(x)}\bra{\phi_y(x)}^{BE}$ and the following extension:
    \begin{align} \label{eq:omega_TXYBE}
        \omega^{TXYBE}:=\Sigma_{t,x,y} R(t|x)p(y|x)p(x) \ketbra{t}^T \otimes \ketbra{x}^X \otimes \ketbra{y}^Y \otimes \ketbra{\phi_y(x)}^{BE}\;. 
    \end{align}
    We arrive at the following chain of equations:
    \begin{align*} \label{eq:key_to_Icoh}
        I(X;B|T)-I(X;E|T)&\overset{a}{=}I(X,Y;B|T)-I(Y;B|X,T)-\left[ I(X,Y;E|T)-I(Y;E|X,T) \right]\\
        &=I(X,Y;B|T)-I(X,Y;E|T)- \left[ I(Y;B|X,T)-I(Y;E|X,T) \right]\\
        & \overset{b}{\leq} I(X,Y;B|T)-I(X,Y;E|T) \numberthis
    \end{align*}
    where (a) follows from chain rule for mutual information; and (b) holds since $I(X,Y;B|T) \geq I(X,Y;E|T)$ by data processing inequality for mutual information as the state $\rho^{ABE}$ is degradable from $B \to E$ and conditioning is done with respect to the classical system.
    
    Now, we need to show that $I(X,Y;B|T)-I(X,Y;E|T) \leq I_c(\rho)$. For this,  the key idea is to consider the coherent version of the state obtained from the application of $\Lambda$ and the channel $R$. We define the following pure state:
    \begin{align} \label{eq:omega}
    \ket{\omega}^{A'TT'XX'BE} &:= V^{A \to A'XX'TT'}\ket{\rho}^{ABE}=\sum_{t,x} \sqrt{R(t|x)}\sqrt{p(x)} \ket{t}^{T'} \otimes \ket{t}^{T} \otimes \ket{x}^{X'} \otimes \ket{x}^{X} \otimes \ket{\rho_x}^{A'BE}
    \end{align}
    where $A' \cong A$, $V^{A \to A'XX'TT'}$ is the isometry that is the composition of the Stinespring isometry of the POVM $\Lambda$ followed by the Stinespring isometry of the channel $R$. Note that $\ket{\omega}^{A'TT'XX'BE}$ is a purification of the state $\omega^{TXBE}$.

    Now consider the following chain of equations with the identification that $A'':= A'XX'$:
    \begin{align*} \label{eq:mutual_info_to_coh}
        I(X,Y;B|T)-I(X,Y;E|T) &= H(B|T)-H(E|T)- \left[ H(XYB|T)-H(XYE|T)\right]\\
        &= H(B|T)-H(E|T) -\left[ H(XYBT)-H(XYET)\right] \\
        &\overset{a}{=} H(B|T)-H(E|T)\\
        &\overset{b}{=} H(BT)-H(A'XX'T'B)\\
        &\overset{c}{=}H(BT)-H(A'XX'TB)\\
        &= I_c(A''>B,T)\\
        &\overset{d}{\leq} D_\rightarrow(\rho)\\
        & \overset{e}{=} I_c(\rho) \numberthis
    \end{align*}
    where (a) follows from the observation that the classical-quantum states on systems $XYBT$ and $XYET$ respectively, have the same eigen values (as can be seen from the state given by Equation~\ref{eq:omega_TXYBE}) and thus $H(XYET)=H(XYBT)$; (b) follows from the state defined in Equation~\ref{eq:omega} and (c) follows since the state on $T$ and $T^\prime$ is the same. The distillable entanglement is given by the optimization of the coherent information over all quantum instruments (Equation~\ref{eq:distil_entanglement}), (d) follows from the construction of the quantum instrument $\mathcal{D}^{A \to A''T}(\ketbra{\rho}^{ABE}):=\Tr_{T'} \left[ V^{A \to A'XX'TT'} \ketbra{\rho}^{ABE} V^\dagger \right]$, where the quantum systems of the instrument output are $A'XX'$ and the classical output register (for LOCC) is $T$;
    (e)  follows from the Fact~\ref{fact:degradable_distil_ebit}. 

    We have thus shown that for any degradable state, its one-way distillable secret key is the same as its coherent information.
\end{proof}
\end{lemma}
\begin{remark}
The above result can also be easily deduced from \cite[Equation~1.9]{Hirche_Key}, where it was shown that for a pure state $\rho^{ABC}$: $D_\rightarrow(\rho^{AB})\leq K_\rightarrow(\rho^{AB}) \leq D_\rightarrow(\rho^{AB})+D_\rightarrow({\rho}^{AC})$ and for a degradable state $\rho^{AB}$, one has that $D_\rightarrow({\rho}^{AC})=0$ from a no-cloning argument, thus implying the result. We give a more direct proof above.
\end{remark}
\end{document}